\documentclass{article}
\usepackage{amsmath}
\usepackage{amssymb}
\usepackage{mathrsfs}
\usepackage{graphicx}
\usepackage{float}
\usepackage{caption}
\usepackage{subcaption}

\usepackage[margin=2cm]{geometry}

\begin{document}

\title{{\bf \Large Violations of the Kerr and Reissner-Nordstr\"om bounds:} \\ {\Large horizon versus asymptotic quantities}}

\vspace{0.5cm}

 \author{
 {\large Jorge F. M. Delgado}\footnote{jorgedelgado@ua.pt}, \
{\large Carlos A. R. Herdeiro}\footnote{herdeiro@ua.pt} \ and
{\large Eugen Radu}\footnote{eugen.radu@ua.pt} 
\\ 
\\
{\small Departamento de F\'\i sica da Universidade de Aveiro and} \\ 
{\small Center for Research and Development in Mathematics and Applications (CIDMA)} \\ 
{\small   Campus de Santiago, 3810-183 Aveiro, Portugal}
}


\date{June 2016}
\maketitle

\begin{abstract}
A central feature of the most elementary rotating black hole (BH) solution in General Relativity is the Kerr bound, which, for vacuum Kerr BHs, can be expressed either in terms of the ADM or the horizon ``charges''. This bound, however, is not a fundamental properties of General Relativity and stationary, asymptotically flat, regular (on and outside an event horizon) BHs are known to violate the Kerr bound, both in terms of their ADM  and horizon quantities. Examples include the recently discovered Kerr BHs with scalar~\cite{Herdeiro:2014goa} or Proca hair~\cite{Herdeiro:2016tmi}.  Here, we point the fact that the Kerr bound in terms of \textit{horizon quantities} is also violated by well-known rotating and charged  solutions, known in closed form, such as the Kerr-Newman and Kerr-Sen BHs. For the former, moreover, we observe that the Reissner-Nordstr\"om (RN) bound is also violated in terms of horizon quantities, even in the static ($i.e$  RN) limit. For the latter, by contrast, the existence of charged matter outside the horizon, allows a curious invariance of the charge to mass ratio, between ADM and horizon quantities. Regardless of the Kerr bound violation, we show that in all case, the event horizon \textit{linear} velocity~\cite{Herdeiro:2015moa} never exceeds the speed of light. Finally, we suggest a new type of informative parameterization for BH spacetimes where part of the asymptotic charges is supported outside the horizon.
\end{abstract}

\section{Introduction}
The current year (2016) marks the 100th anniversary of the Schwarzschild solution~\cite{Schwarzschild:1916uq}. This is the simplest exact  black hole (BH) solution of General Relativity, and it is parameterized solely by its mass, $M$. Adding electric charge, $Q$, or angular momentum, $J$, to the Schwarzschild solution yields the Reissner-Nordstr\"om (RN) or Kerr spacetimes, respectively (see $e.g.$~\cite{PoissonRelativistsToolkit,TownsendBlackHoles}). In either case, these elementary solutions have a bound in their parameters space. There is a maximum amount of charge or angular momentum for a given mass (in units where $G=c=1=4\pi \epsilon_0$, to be used throughout):
\begin{equation}
q_E\equiv \frac{|Q|}{M}\leqslant 1 \ , \ \ \ {\rm for \ RN} \ , \qquad j\equiv \frac{|J|}{M^2}\leqslant 1 \ ,  \ \ \ {\rm for \ Kerr} \ ;
\label{bounds}
\end{equation}
when such bounds are violated, the spacetime describes a naked singularity rather than a BH. 

A natural question is:  ``how fundamental are these bounds in BH physics?"  
(see~\cite{Herdeiro:2015moa} for a recent related discussion about the Kerr bound). 
On the one hand, it is well known that these bounds are violated by physical objects outside the scope of strong gravity. For instance, the RN bound is violated by an electron and the Kerr bound is violated by any rotating macroscopic object, in both cases by many orders of magnitude. Within the scope of strong gravity, on the other hand, it is reasonable to expect that gravitational collapse can only yield a BH if the dominant interaction is attractive; thus, too much (one sided) electric charge  or rotation has the potential to prevent such collapse into a BH. This reasoning therefore suggests that the total charge or angular momentum should be bounded by the total energy/mass in a BH spacetime, a suggestion that is vindicated by the bounds~\eqref{bounds} found at the level of the exact solutions.

In this letter we point out that the situation can actually be more interesting in terms of \textit{horizon quantities}, which are reviewed in Section~\ref{sec_quantities}. Our motivation comes from the recently discovered Kerr BHs with scalar~\cite{Herdeiro:2014goa,Herdeiro:2015gia} or Proca~\cite{Herdeiro:2016tmi} hair, wherein part of the spacetime energy and angular momentum is stored outside the horizon in a ``matter" field, and the Kerr bound can be violated at the level of both ADM and horizon ``charges".\footnote{Since angular momentum is not defined in the ADM formalism, by ADM angular momentum we mean the angular momentum computed at spatial infinity as a Komar integral.} This situation will be reviewed in Section~\ref{sec_hairy},  where we consider an informative parameterization for these spacetimes to make explicit the division of conserved charges between the horizon and the exterior region. Motivated by the violation of the Kerr bound in terms of horizon quantities that occurs in these examples, we ask whether this is due to the ``hairiness" of these spacetimes, $i.e$ due to the matter that exists outside the horizon in equilibrium with the BH, and which is associated to an independent, locally conserved charge, but not to a Gauss law (primary hair).

To tackle this question, we examine, in Section~\ref{sec_charged} two families of rotating BHs known in closed form, where part of the spacetime energy and angular momentum  is also stored outside the horizon. These are the well known Kerr-Newman~\cite{Newman:1965my} and Kerr-Sen~\cite{KerrSen} BHs. The former is ``bald"; the latter can be faced as possessing a type of secondary hair (see $e.g.$ the review~\cite{Herdeiro:2015waa}), since there is a scalar field outside the horizon, but which is not associated to an independent scalar charge. We observe that, despite the fact that the Kerr bound is never violated in terms of ADM quantities for these solutions, it is indeed violated in terms of horizon quantities. Moreover, for the Kerr-Newman family, also the RN bound is violated in terms of horizon quantities, even in the static limit. Thus we face the curious situation that  the RN bound is violated by RN BHs, in terms of horizon quantities. For the Kerr-Sen case, by contrast, the RN bound is not violated and we point out a curious invariance of the charge to mass ratio. Finally we discuss that, albeit the Kerr bound is violated, all examples discussed here obey a different bound on the rotation: the linear horizon velocity bound proposed in~\cite{Herdeiro:2015moa}. Some final remarks are made in Section~\ref{sec_discussion}.

\section{Horizon $vs.$ asymptotic quantities}
\label{sec_quantities}
In a stationary and axi-symmetric spacetime, the Killing symmetries allow a quasi-local definition of energy, $M_S$ and angular momentum, $J_S$,  in a volume bounded by  a co-dimension two spacelike surface $S$. These quantities are given in terms of Komar integrals, which in four spacetime dimensions take the form~\cite{PoissonRelativistsToolkit,TownsendBlackHoles}:
\begin{equation}
M_S = - \frac{1}{8\pi} \oint_S dS_{\mu \nu} D^\mu t^\nu\ , \qquad J_S = \frac{1}{16\pi} \oint_S dS_{\mu \nu} D^\mu \phi^\nu\ ,
\end{equation}
where $t^\nu$ and $\phi^\nu$ are the timelike and rotational Killing vectors, respectively, and $dS_{\mu \nu}$ is the surface element,
\begin{equation}
dS_{\mu \nu} = 2 n_{[\mu} l_{\nu]} \sqrt{-g} d\Omega  \ ,
\end{equation}
and $n_\mu$ and $l_\nu$ are the timelike and spacelike vectors normal to the surface.

Taking $S$ as a round  2-sphere at spatial infinity, $S^2_\infty$, for an asymptotically flat spacetime, these quantities become the ADM ones $(M_S,J_S)\rightarrow (M,J)$; on the other hand, taking $S$ as a spatial section of the event horizon, $\mathscr{H}$, these quantities become the horizon mass and angular momentum $(M_S,J_S)\rightarrow (M_H,J_H)$. An application of Gauss' theorem relates these two sets of quantities~\cite{PoissonRelativistsToolkit}: 
\begin{eqnarray}\label{TotalMass}
M = M_H -2\int_\Sigma dS_\mu\bigg( T_{\nu}^{\ \mu}  t^{\nu}-\frac{1}{2}T t^\mu \bigg),~~~
 J = J_H + \int_\Sigma dS_\mu \left( T_{\nu}^{\ \mu}  \phi^{\nu} -\frac{1}{2}T  \phi^{\mu} \right) \ ,
\end{eqnarray}
where $\Sigma$ is a spacelike surface, bounded by $S^2_\infty$ and  $\mathscr{H}$.
For Ricci flat spacetimes, $M=M_H$ and $J=J_H$; for a non-vacuum BH spacetime, in general, these quantities differ. To measure this difference, we have found useful to define
\begin{eqnarray}
p \equiv \frac{M_H}{M}~~~{\rm and}~~~ q\equiv \frac{J_H}{J}\ ,
\end{eqnarray}
corresponding to the fractions of ADM mass and angular momentum which are stored inside the
horizon.

The electric charge can be computed by the covariant form of Gauss' law:
\begin{equation}\label{BlackHoleCharge}
Q_S = \frac{1}{8\pi} \oint_S dS_{\mu \nu} F^{\mu \nu} \ .
\end{equation}
The asymptotic charge, $Q$, is obtained by taking $S=S^2_\infty$, whereas the horizon charge, $Q_H$, is obtained by taking $S=\mathscr{H}$. These two quantities are related by applying again Gauss' theorem:
\begin{equation}
Q=Q_H+\frac{1}{4\pi}\int_\Sigma dS_\mu D_\nu F^{\mu\nu} \ ,
\end{equation}
Thus, as expected, the two charges may differ when there are sources to Maxwell's equations outside the horizon.

\section{Numerical spacetimes: hairy Kerr-BHs}
\label{sec_hairy}
\begin{figure}[H]
\centering
\begin{subfigure}[b]{0.45\textwidth}
\includegraphics[width=1.1\linewidth]{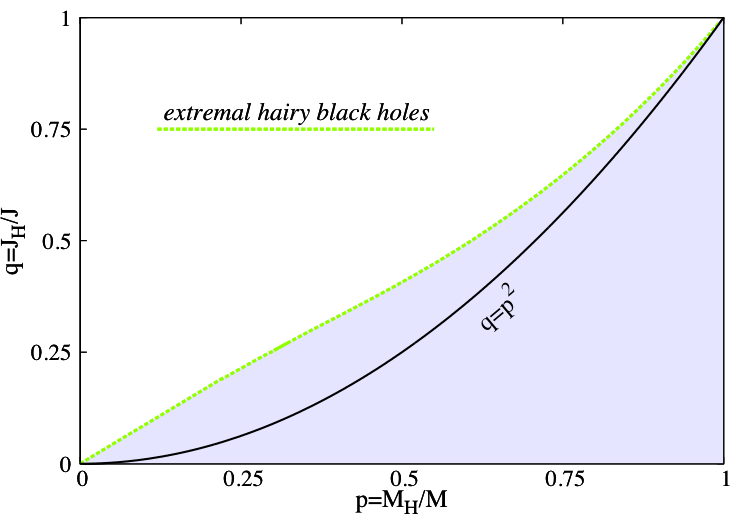}
\caption{}
\label{figures1}
\end{subfigure}
\ \ \ \ \ \ 
\begin{subfigure}[b]{0.45\textwidth}
\includegraphics[width=1.1\linewidth]{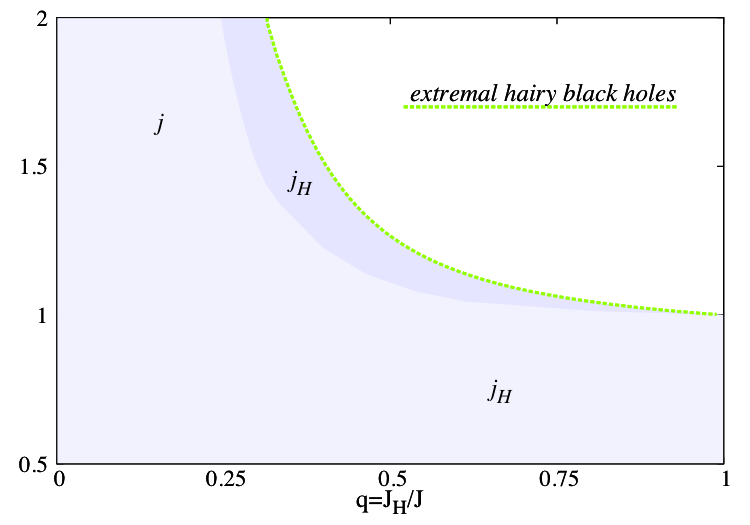}
\caption{}
\label{figures2}
\end{subfigure}
\caption{ {\small (a) Fraction of angular momentum in the horizon, $q$, $vs.$ fraction of the mass in the horizon, $p$. Solution above the black solid line have a horizon dimensionless angular momentum larger than the ADM one.  (b) Dimensionless ADM, $j$ (light shaded area), and horizon, $j_H$ (light plus dark shaded area) angular momentum, $vs.$ fraction of the angular momentum in the horizon, $q$. In both panels, the dotted green line corresponds to extremal Kerr BHs with scalar hair.}}
\label{figure0}
\end{figure}

We consider Einstein's gravity minimally coupled to a massive complex scalar field, $\Psi$ or to a massive complex vector ($i.e$ Proca) field, $\mathcal{A}$. The action, including the two types of matter fields, is:
\begin{equation}
S=\frac{1}{4}\int  d^4x \sqrt{-g} \bigg[R
-2\left(
          \partial_ \alpha \bar{\Psi}  \partial^\alpha \Psi  + \partial_ \alpha {\Psi}  \partial^\alpha \bar{\Psi} 
	+2 \mu^2 \bar{\Psi}\Psi 
	\right) 
	- \left( \mathcal{F}_{\alpha\beta}\bar{\mathcal{F}}^{\alpha\beta}
+2\mu^2\mathcal{A}_\alpha\bar{\mathcal{A}}^\alpha 
    \right) 
 \bigg]  \ ,
\end{equation}
where $\mathcal{F}=d\mathcal{A}$, $\mu$ is the mass of both the scalar and the Proca field  and an overbar denotes complex conjugate. Non-extremal Kerr BHs with scalar~\cite{Herdeiro:2014goa}  or Proca~\cite{Herdeiro:2016tmi}  
hair have been found  with the ansatz\footnote{ Note that the Kerr metric 
can be written in this form \cite{Herdeiro:2015gia}.}:
\begin{equation}
ds^2=-e^{2F_0(r,\theta)}N(r)dt^2+e^{2F_1(r,\theta)}\left[\frac{dr^2}{N(r)}+r^2d\theta^2\right]+r^2\sin^2\theta e^{2F_2(r,\theta)}[d\varphi-W(r,\theta)dt]^2 \ ,
\label{ansatzkbhssh}
\end{equation}
where $N(r)\equiv 1-r_H/r$ and $r_H$  is a constant, the radial coordinate of the event horizon.  The ``matter" ansatz, on the other hand, is
\begin{equation}
\Psi(t,r,\theta,\varphi)=e^{-i\omega t+im\varphi}\phi(r,\theta) \ ,
\label{ansatzscalar}
\end{equation}
for the BHs with scalar hair~\cite{Herdeiro:2014goa} and
\begin{equation}
A(t,r,\theta,\varphi)=e^{-i\omega t+im\varphi}\{i[V(r,\theta)dt+H_3(r,\theta)\sin\theta d\varphi]+H_1(r,\theta) dr+H_2(r,\theta) d\theta\} \ ,
\label{ansatzproca}
\end{equation}
for the BHs with Proca hair~\cite{Herdeiro:2016tmi}. The two constants $\omega,m$ are the frequency and azimuthal quantum number, $\omega\in \mathbb{R^+}$, $m\in\mathbb{Z}/\{0\}$. The four unknown metric functions plus the scalar profile function/the four Proca potential  functions solve a set of nonlinear partial differential equations and are determined numerically~\cite{Herdeiro:2015gia,Herdeiro:2016tmi}.\footnote{So far, no BHs with \textit{both} scalar and Proca hair have been found, although these are likely to exist. The previously mentioned solutions have either scalar or Proca hair, not both.} In Fig.~\ref{figures1} we display the distribution of Kerr BHs with scalar hair in a $q\equiv J_H/J$ $vs.$ $p\equiv M_H/M$ diagram (shaded blue region). 
The top right corner is a degenerate limit corresponding to the vacuum limit (all corresponding Kerr solutions), whereas the bottom left corner corresponds to (the also degenerate) solitonic limit (scalar boson stars~\cite{Schunck:2003kk}). The black solid line corresponds to $q=p^2$ and one can see that there are hairy BHs above it, corresponding to solution that have dimensionless horizon angular momentum, $j_H\equiv J_H/M_H^2$, greater than the dimensionless ADM angular momentum, $j\equiv J/M^2$. In Fig.~\ref{figures2} we exhibit $j,j_H$ $vs.$ $q$ diagram, where one can clearly see that there are solutions violating the Kerr bound both in terms of ADM and horizon quantities. For Kerr BHs with Proca hair the corresponding diagrams are qualitatively similar; as such we will not present them here.

\section{Closed form solutions: charged Kerr-BHs}
\label{sec_charged}

\subsection{Kerr-Newman}

The Kerr-Newman BH~\cite{Newman:1965my} is a solution to the Einstein-Maxwell theory, described by the action:
\begin{equation}
\label{KerrNewmanAction}
\mathcal{S} =\frac{1}{4} \int d^4 x \sqrt{-g} \left(  R -  F_{\mu\nu} F^{\mu\nu} \right) \ ,
\end{equation}
where $R$ is the Ricci scalar, the Maxwell field strength is $F_{\mu\nu} \equiv \partial_\mu A_\nu - \partial_\nu A_\mu$ and $A_\mu$ is the Maxwell 4-potential. The corresponding Einstein-Maxwell equations are,
\begin{eqnarray}
 R_{\mu\nu} - \frac{1}{2} g_{\mu\nu} R = 2 \left( F_{\mu\alpha} {F_\nu}^\alpha - \frac{1}{4} g_{\mu \nu} F_{\alpha \beta} F^{\alpha \beta} \right) \ , \qquad  D_\nu {F^\nu}_{\mu} = 0 \ .
 \label{eqmotnewman}
\end{eqnarray}
The Kerr-Newman configuration describes a BH with ADM mass $M$, ADM angular momentum per unit mass $a=J/M$ and electric charge $Q$, and in Boyer-Lindquist coordinates~\cite{Boyer:1966qh} $(t,r,\theta, \phi)$ it reads,
\begin{equation}\label{KerrNewmanMetric}
ds^2 = - \frac{\Delta}{\Sigma} \left( dt -a \sin^2 \theta d\phi \right)^2 +  \Sigma \left( \frac{dr^2}{\Delta} + d\theta^2 \right) + \frac{\sin^2 \theta}{\Sigma} \left[ a dt - \left( \Sigma+a^2\sin^2\theta \right) d\phi \right]^2 \ ,
\end{equation}
and
\begin{equation}\label{ElectromagneticVectorKerrNewman}
A_\mu dx^\mu = -\frac{Q r}{\Sigma} \left( dt - a \sin^2 \theta d\phi \right) \ ,
\end{equation}
where 
\begin{equation}
\Delta \equiv r^2 - 2Mr + a^2 + Q^2 \ , \qquad  \Sigma \equiv r^2 + a^2 \cos^2 \theta \ .
\label{deltasigma}
\end{equation}

To address horizon quantities, we must use a regular coordinate system on the horizon, which is not the case of Boyer-Lindquist coordinates. Thus, we introduce the following coordinate transformation,
\begin{equation}
dv = dt + \frac{\Sigma + a^2\sin^2\theta}{\Delta}dr \hspace{10pt} \text{and} \hspace{10pt} d\psi = d\phi + \frac{a}{\Delta} dr \ ,
\label{ctkn}
\end{equation}
which yields the Kerr-Newman metric in ingoing Eddington-Finkelstein-type coordinates
\begin{equation}
ds^2 = -\frac{\Delta}{\Sigma} \left(dv - a \sin^2 \theta d \psi \right)^2 + 2 dr \left(dv - a \sin^2 \theta d\psi \right) + \Sigma d\theta^2 + \frac{\sin^2 \theta}{\Sigma} \left[a dv - (\Sigma+a^2\sin^2\theta) d\psi \right]^2 \ .
\label{lekn2}
\end{equation}

\subsection{Kerr-Sen}
The Kerr-Sen BH~\cite{KerrSen} is a solution to the low energy effective field theory of the heterotic string,  compactified on a 6-torus. In the string frame, the corresponding four dimensional action is
\begin{equation}
\mathcal{S} = \frac{1}{4} \int d^4 x \sqrt{-G} e^{-\Phi} 
\left( R 
- \frac{1}{12} H_{\mu \nu \rho} H^{\mu \nu \rho} 
+ G^{\mu \nu} \partial_\mu \Phi \partial_\nu \Phi 
- \frac{1}{8} F_{\mu\nu} F^{\mu\nu} 
\right) \ ,
\end{equation}
where $\Phi$ is the dilation field, the Neveu-Schwarz field strength 3-form $H_{\mu \nu \rho}$ is defined as $ H_{\mu \nu \rho} =3 \partial_{[\mu} B_{\nu \rho]} - 3 A_{[\mu} F_{\nu\rho]}/4$, in terms of the Neveu-Schwarz 2-form potential, $B_{\mu\nu}$, 
and the Maxwell field strength, $F_{\mu\nu}$ and potential, $A_\mu$. $G_{\mu\nu}$ is the string-frame metric; it is related to the Einstein-frame metric $g_{\mu\nu}$ as $g_{\mu\nu} = e^{-\Phi} G_{\mu\nu}$. Performing this conformal transformation, together with the rescalings $\Phi \rightarrow 2\Phi$ and $A_\mu \rightarrow 2\sqrt{2} A_\mu$~\cite{WuKerrSen}, one obtains the following Einstein-frame action,
\begin{equation}\label{KerrSenAction}
\mathcal{S} = \frac{1}{4} \int d^4 x \sqrt{-g} \left( R - 2 \partial_\mu \Phi \partial^\mu \Phi - e^{-2\Phi} F_{\mu\nu}F^{\mu\nu} - \frac{1}{12} e^{-4\Phi} H_{\mu\nu\rho} H^{\mu\nu\rho} \right) \ .
\end{equation} 
In terms of the rescaled fields, the 3-form $H_{\mu\nu\rho}$ becomes $H_{\mu\nu\rho} = 3\partial_{[\mu} B_{\nu \rho]} - 6  A_{[\mu} F_{\nu\rho]}$. The equations of motion obtained in the Einstein-frame are:
\begin{eqnarray}
 \Box \Phi = - \frac{1}{2} e^{-2 \Phi} F_{\mu\nu}F^{\mu\nu} - \frac{1}{12} e^{-4\Phi} H_{\mu\nu\rho}H^{\mu\nu\rho} \ ,  \qquad 
D_{\rho} \left( e^{-2\Phi} {F^\rho}_\mu  \right) = -\frac{1}{2}e^{-4\Phi} H_{\mu\nu\rho} F^{\nu\rho} \ , 
 \label{eqmotsen}
 \end{eqnarray}
 \begin{eqnarray}
D_\alpha \left( e^{-4\Phi} {H^\alpha}_{\mu\nu} \right) = 0 \ , \qquad 
 R_{\mu\nu} - \frac{1}{2} g_{\mu\nu} R = T^{(DF)}_{\mu\nu} + T^{(EM)}_{\mu\nu} + T^{(NS)}_{\mu\nu} \ , 
\end{eqnarray}
where the dilaton field, electromagnetic and Neveu-Schwarz field energy-momentum tensors are, 
\begin{equation} 
T^{(DF)}_{\mu\nu} = 2 \left(\partial_\mu \Phi \partial_\nu \Phi - \frac{1}{2}g_{\mu\nu} \partial_\rho \Phi \partial^\rho \Phi\right) \ , \qquad T^{(EM)}_{\mu\nu} = 2 e^{-2\Phi}\left( F_{\mu\rho} {F_\nu}^\rho - \frac{1}{4} g_{\mu\nu}  F_{\rho\sigma} F^{\rho\sigma}\right) \ ,
 \end{equation}
 \begin{equation}
\nonumber
T^{(NS)}_{\mu\nu} = \frac{1}{4} e^{-4\Phi} \left(H_{\mu\rho\sigma} {H_\nu}^{\rho\sigma} - \frac{1}{6} g_{\mu\nu}  H_{\rho\sigma\lambda} H^{\rho\sigma\lambda}\right) \ .
\end{equation} 
Observe, in particular, from the second equation in~\eqref{eqmotsen}, that the Neveu-Schwarz field (and the dilaton) act as sources for the Maxwell field. This contrast to the Einstein-Maxwell theory above, where there are no sources for the Maxwell field, as can be seen from the Maxwell equations in~\eqref{eqmotnewman}.

The Kerr-Sen configuration solves these equations of motion, and it is given by, in Boyer-Lindquist coordinates $(t,r,\theta,\phi)$, \cite{WuKerrSen}
\begin{equation}\label{KerrSenMetric}
ds^2 = - \frac{\Delta'}{\Sigma'} \left( dt -a \sin^2 \theta d\phi \right)^2 +\Sigma'  \left( \frac{dr^2}{\Delta'} + d\theta^2 \right) + \frac{\sin^2 \theta}{\Sigma'} \left[a dt - \left( \Sigma' + a^2 \sin^2 \theta \right) d\phi \right]^2  \ ,
\end{equation}
\begin{equation}\label{ElectromagneticVectorKerrSen}
A_\mu dx^\mu = -\frac{Q r}{\Sigma'} \left( dt - a \sin^2 \theta d\phi \right) \ , \qquad \Phi = - \frac{1}{2} \ln \left( \frac{\Sigma'}{\Sigma' - b r} \right) \ , \qquad 
B_{\mu\nu}dx^\mu \wedge dx^\nu = 2a \sin^2 \theta \ \frac{ b r}{\Sigma'}dt\wedge d\phi  \ ,
\end{equation}
where 
\begin{equation}
\Delta' \equiv  r^2 - 2Mr + a^2 + br=\Delta -Q^2 +br \ , \qquad   \Sigma' \equiv r^2 + a^2 \cos^2 \theta + br=\Sigma +br \ , \qquad  b \equiv  \frac{Q^2}{M} \ , 
\end{equation}
where $Q$ is the electric charge measured at infinity and $M$ is the ADM mass~\cite{WuKerrSen}.

Again, a regular coordinate system on the horizon is achieved by the following coordinate transformation,
\begin{equation}
dv = dt + \frac{\Sigma'+a^2\sin^2\theta}{\Delta'} dr \hspace{10pt} \text{and} \hspace{10pt} d\psi = d\phi + \frac{a}{\Delta'} \ ,
\label{ctks}
\end{equation}
which yields the line element
\begin{equation}
ds^2 = - \frac{\Delta'}{\Sigma'} \left( dv -a \sin^2 \theta d\psi \right)^2 + 2 dr \left( dv - a \sin^2 \theta d\psi \right) + \Sigma' d\theta^2  + \frac{\sin^2 \theta}{\Sigma'} \left[a dv- \left( \Sigma' + a^2 \sin^2 \theta \right) d\psi \right]^2 \ .
\label{leks2}
\end{equation}
We remark that the line elements of the Kerr-Newman metric \eqref{KerrNewmanMetric} and \eqref{lekn2}, as well as their coordinate transformation~\eqref{ctkn} and the gauge potential~\eqref{ElectromagneticVectorKerrNewman}, are interchanged with those of the Kerr-Sen metric,  eqs.~\eqref{KerrSenMetric} and \eqref{leks2}, and the corresponding coordinate transformation~\eqref{ctks} and 1-form gauge potential~\eqref{ElectromagneticVectorKerrSen}, by interchanging (see $e.g.$~\cite{HerdeiroKerrSen})
\begin{equation}
(\Sigma, \Delta) \ \  \longleftrightarrow \ \ (\Sigma', \Delta')=(\Sigma+br,\Delta-Q^2+br) \ .
\end{equation}

\subsection{Horizon and asymptotic quantities}
We now specialize the physical quantities of Section~\ref{sec_quantities} to the Kerr-Newman and Kerr-Sen solutions. To compute the horizon quantities,  the timelike and spacelike vectors are
\begin{equation}\label{nullvectors}
n_\mu dx^\mu = \left( 1- \frac{a^2 \sin^2 \theta}{\Sigma + a^2\sin^2\theta} \right) dr  \hspace{10pt} ; \hspace{10pt} l_\nu dx^\nu = - dv + \frac{a^2 \sin^2 \theta}{2(\Sigma + a^2\sin^2\theta)} dr \ , 
\end{equation}
for Kerr-Newman, and the same expression with $\Sigma\rightarrow \Sigma'$ for Kerr-Sen. Then
\begin{equation}
M_{H} = \left\{-\frac{M(\Sigma+a^2\sin^2\theta)}{2} \int_0^\pi  \frac{\Sigma' -  2r^2}{\Sigma^2} \sin \theta d\theta\right\}_{\rm r=r_H} \ ,
\end{equation}
for Kerr-Newman, and the same expression for Kerr-Sen \textit{interchanging} $\Sigma\leftrightarrow \Sigma'$. Performing the integral and using 
\begin{equation}Êr_H = M + \sqrt{M^2-Q^2-\frac{J^2}{M^2}} \ , 
\label{rhkn}
\end{equation} 
for Kerr-Newman and 
\begin{equation}
r_H = M - \frac{Q^2}{2M} + \sqrt{\left(M-\frac{Q^2}{2M}\right)^2- \frac{J^2}{M^2}}  \ ,
\label{rhks}
\end{equation} 
for Kerr-Sen, one obtains for these two cases, respectively:
\begin{equation}\label{MassHorizon}
M_{H} =M\left(1-\frac{Q^2}{2Mr_H}\right) \left[ 1 - \frac{Q^2}{a r_H} \arctan \left( \frac{a}{r_H} \right) \right] \ , \qquad {\rm Kerr-Newman} \ ,
\end{equation}
\begin{equation}\label{MassHorizonKerrSen}
M_H =  M \left[ \frac{r_H^2+br_H/2}{ r_H^2+b r_H } - \frac{ Q^2 r_H^2}{a (b r_H + r_H^2)^{3/2}}  \arctan \left(\frac{a}{\sqrt{br_H + r_H^2}}\right) \right] \ , \qquad {\rm Kerr-Sen} \ .
\end{equation}

Proceeding similarly for the horizon angular momentum, one arrives at
\begin{equation}\label{AngularMomentumHorizon}
J_{H} = J\left(1-\frac{Q^2}{2Mr_H}\right)  \left\lbrace 1 + \frac{Q^2}{2 a^2} \left[ 1 - \frac{a^2 + r_H^2}{a r_H} \arctan \left( \frac{a}{r_H} \right) \right]  \right\rbrace \ , \qquad {\rm Kerr-Newman} \ ,
\end{equation}
\begin{equation}\label{AngularMomentumHorizonKerrSen}
J_H = J \left[ \frac{r_H^2+3br_H/4}{ r_H^2+b r_H } + \frac{b r_H}{4a^2}  - \frac{ Q^2 M r_H^3}{a^3 (b r_H + r_H^2)^{3/2}} \arctan \left( \frac{a}{\sqrt{b r_H + r_H^2}} \right) \right] \ , \qquad {\rm Kerr-Sen} \ .
\end{equation}

Finally, for the horizon charge, one finds:
\begin{equation}\label{ChargeHorizon}
Q_H = Q \ , \qquad {\rm Kerr-Newman} \ ,
\end{equation}
\begin{equation}\label{ChargeHorizonKerrSen}
Q_H =  Q \left[ \frac{r_H^2+br_H/2}{ r_H^2+b r_H }  - \frac{ Q^2 r_H^2}{a(br_H + r_H^2)^{3/2}} \arctan \left( \frac{a}{\sqrt{br_H + r_H^2}} \right)\right] \ , \qquad {\rm Kerr-Sen} \ .
\end{equation}

Using~\eqref{rhkn} and~\eqref{rhks}, together with $a=J/M$ and $b=Q^2/M$, all these quantities can be expressed in terms of $(M,J,Q)$ only, for either case. 

Observe that, for both solutions, 
\begin{equation*}
\lim_{Q \rightarrow 0} M_{H} = M \ , \qquad \lim_{Q \rightarrow 0} J_{H} = J \ , \qquad  \lim_{Q \rightarrow 0} Q_{H} = 0 \ .
\end{equation*}
as expected. As another consistency check, we verify Smarr's formula~\cite{Smarr:1972kt}, written solely in terms of horizon quantities~\cite{TownsendBlackHoles}, that is obeyed by both Kerr-Newman and Kerr-Sen BHs:
\begin{equation}\label{SmarrFormula}
M_H - 2 \Omega_H J_H = \frac{\kappa A_H}{4 \pi} \ ,
\end{equation}
where, $\Omega_H$ is the angular velocity of the horizon, $\kappa$ is the surface gravity and $A_H$ is the area of the spatial sections of the event horizon. For the Kerr-Newman solution 
\begin{align*}
\Omega_H = \frac{a}{r_H^2 + a^2} \ ,  \qquad \kappa = \frac{r_H-M}{r_H^2 + a^2}  \ , \qquad 
A_H= 4 \pi  \left(r_H^2 + a^2\right) \ .
\end{align*}
With these results and equations \eqref{MassHorizon} and \eqref{AngularMomentumHorizon} it is easy to prove that both sides of Smarr's Formula \eqref{SmarrFormula} give $\sqrt{M^2 - a^2 - Q^2}$. For the Kerr-Sen solution, 
\begin{align*}
\Omega_H = \frac{a}{r_H^2 + a^2 + b r} \ , \qquad
\kappa  = \frac{r_H^2 - a^2}{4 M r_H}  \ , \qquad 
A &= 4 \pi  \left(r_H^2 + a^2 + br\right) \ .
\end{align*}
These results, together with \eqref{MassHorizonKerrSen} and \eqref{AngularMomentumHorizonKerrSen} show that both sides of Smarr's Formula \eqref{SmarrFormula} give $\sqrt{(M + b/2)^2 - a^2}$.

\subsubsection{Analysis of horizon quantities}
Let us now analyze the results obtained in the previous section. Starting with the familiar Kerr-Newman case, a curious feature is obtained even in its static limit (RN). Then,~\eqref{MassHorizon} reduces to 
\begin{equation}
M_H=M-\frac{Q^2}{r_H}\ .
\end{equation}

\begin{figure}[H]
\centering

\begin{subfigure}[b]{0.45\textwidth}
\includegraphics[width=1.0\linewidth]{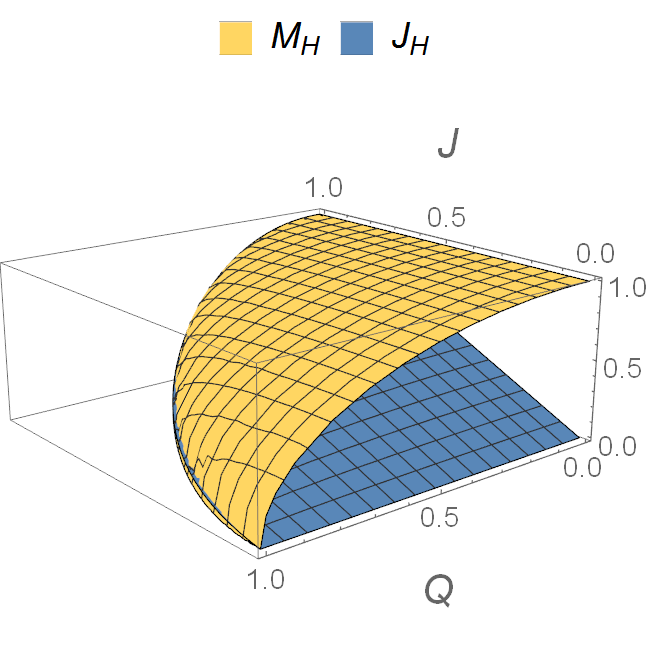}
\caption{}
\label{figurekn1}
\end{subfigure}
\hfill
\begin{subfigure}[b]{0.45\textwidth}
\includegraphics[width=1.0\linewidth]{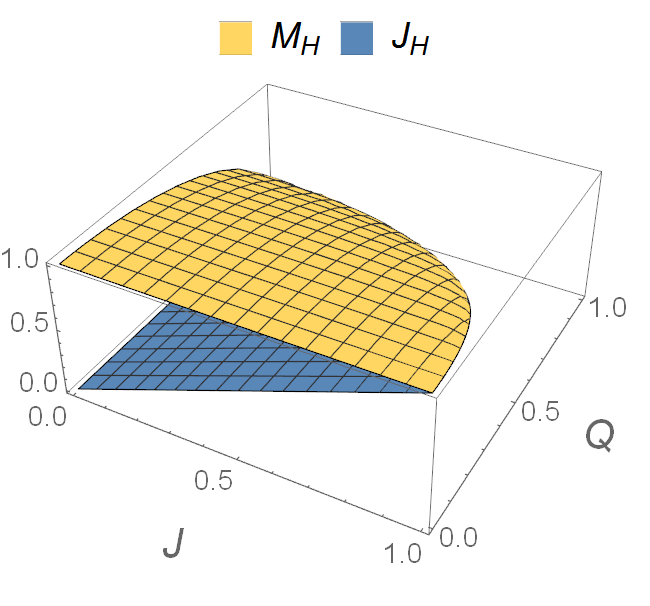}
\caption{}
\label{figurekn2}
\end{subfigure}
\caption{
{\small
Horizon mass and angular momentum for a Kerr-Newman BH in terms of its asymptotic angular momentum and charge.}
}
\label{figure1}
\end{figure}

\begin{figure}[H]
\centering

\begin{subfigure}[b]{0.53\textwidth}
\includegraphics[width=1.0\linewidth]{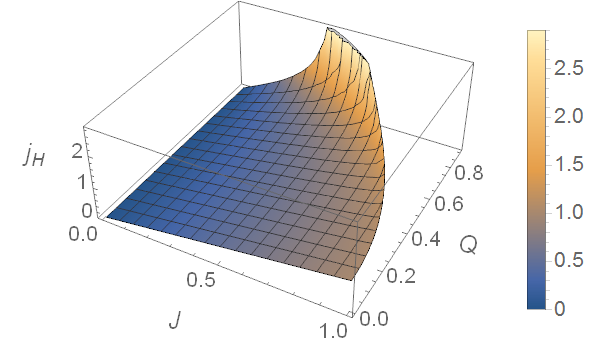}
\caption{}
\label{figurekn3}
\end{subfigure}
\hfill
\begin{subfigure}[b]{0.38\textwidth}
\includegraphics[width=1.0\linewidth]{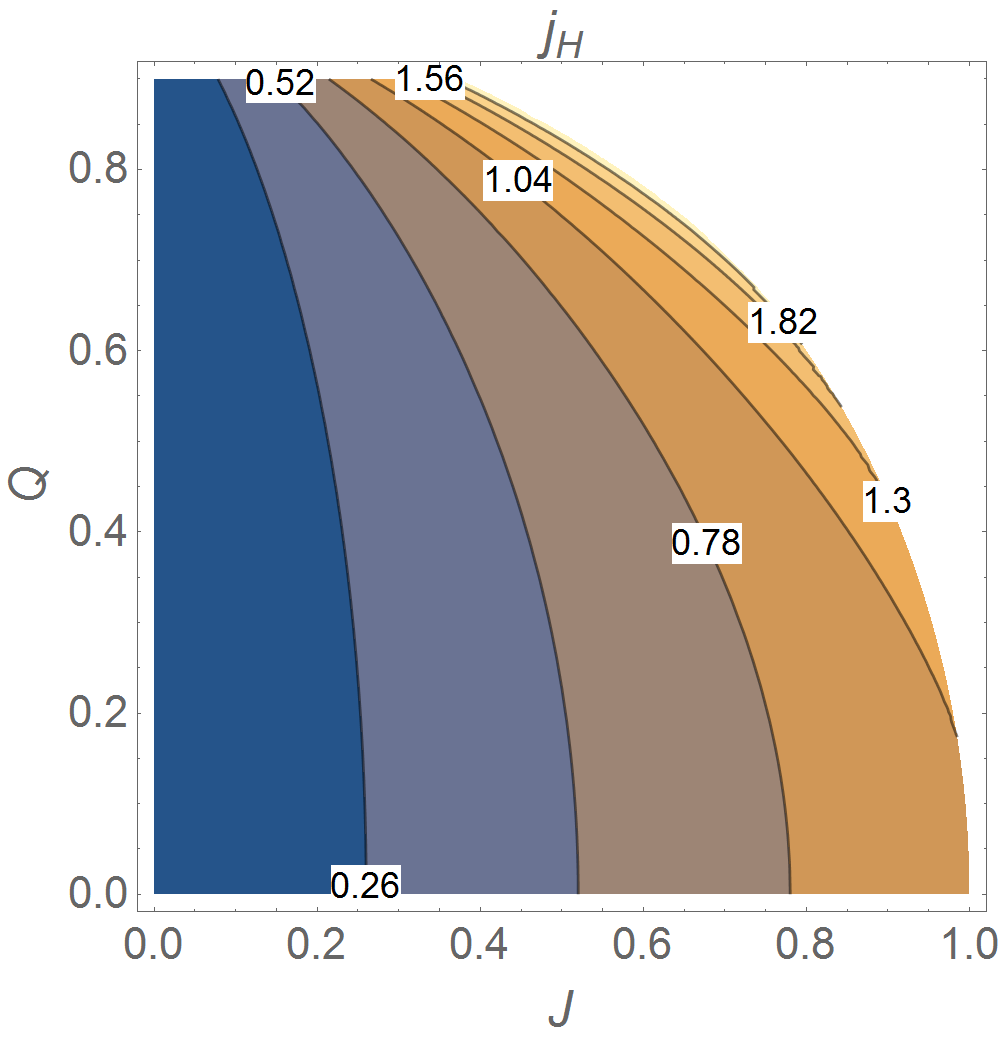}
\caption{}
\label{figurekn4}
\end{subfigure}
\caption{
{\small
 (a) 3D and (b) 2D contour plots of $j_H\equiv J_{H}/M_{H}^2$ for the Kerr-Newman BH. This quantity becomes larger than unity for sufficiently large charge and angular momentum. }
}
\label{figure2}
\end{figure}

The horizon mass, $M_H$, is always smaller than the ADM mass, $M$, as expected from the fact that the electric field outside the horizon carries energy. The former decreases monotonically from $M_H=M$ for an uncharged BH to $M_H=0$, for an extremal BH. This behaviour is clearly seen in Fig.~\ref{figurekn1}

Since for the RN BH there is no charge outside the horizon, $Q_H=Q$ and an immediate consequence is that the dimensionless quantity 
\begin{equation}
q_E^{(H)}\equiv \frac{|Q_H|}{M_H}=\frac{q_E}{\sqrt{1-q_E^2}} \ ,
\end{equation} 
becomes larger than unity. Actually it diverges as extremality is approached. In other words, the Reissner-Nordstr\"om bound is violated in the Reissner-Nordstr\"om solution, in terms of horizon quantities.  

To analyze the more general Kerr-Newman family we fix $M=1$ and vary $Q,J$. Then:
\begin{description}
\item[i)]  for fixed, but non-zero, $J$, the horizon mass $M_H$ decreases with increasing $Q$ -- similarly to the case with $J=0$ -- because a part of the energy is transferred into the electric field outside the horizon - Fig.~\ref{figurekn1}. The horizon angular momentum also decreases: the growing outside \textit{electromagnetic} field carries a larger fraction of the total angular momentum; 
\item[ii)] for fixed charge $Q$, increasing the angular momentum $J$, the horizon angular momentum naturally also increases, but the horizon mass decreases.
\end{description}

\begin{figure}[H]
\centering

\begin{subfigure}[b]{0.45\textwidth}
\includegraphics[width=1.0\linewidth]{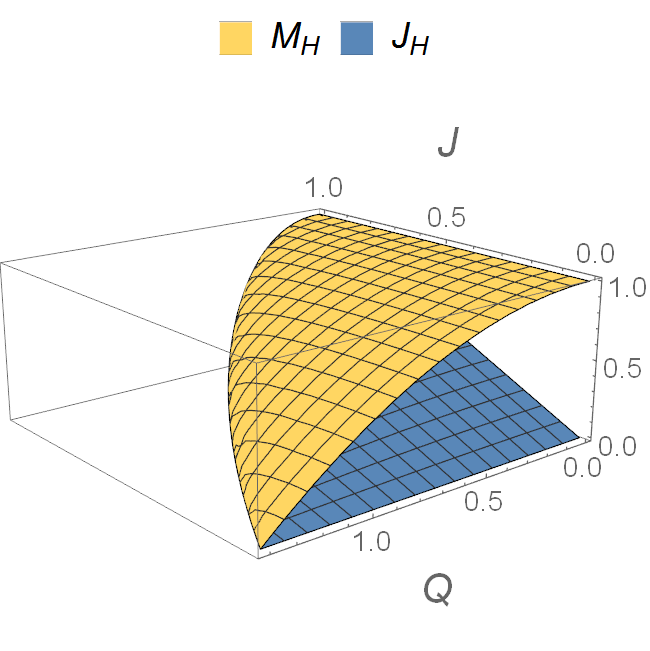}
\caption{}
\label{figureks1}
\end{subfigure}
\hfill
\begin{subfigure}[b]{0.45\textwidth}
\includegraphics[width=1.0\linewidth]{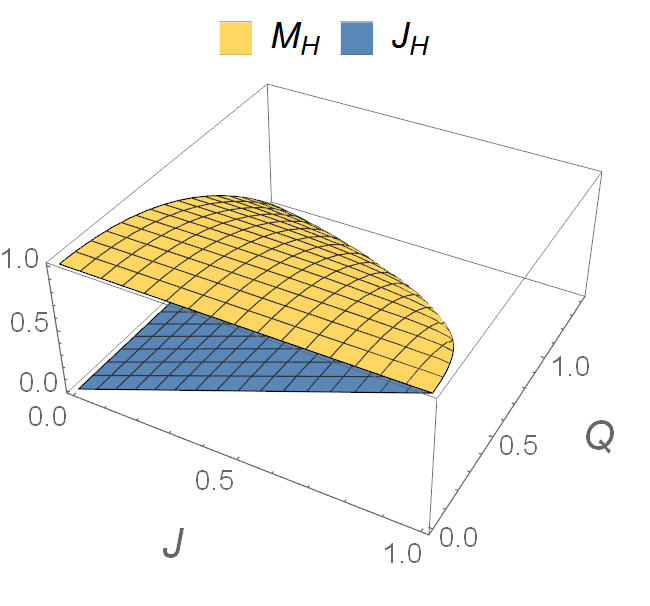}
\caption{}
\label{figureks2}
\end{subfigure}
\caption{
{\small
Horizon mass and angular momentum for a Kerr-Sen BH in terms of its asymptotic angular momentum and charge.}
}
\label{figure3}
\end{figure}
 
\begin{figure}[H]
\centering

\begin{subfigure}[b]{0.53\textwidth}
\includegraphics[width=1.0\linewidth]{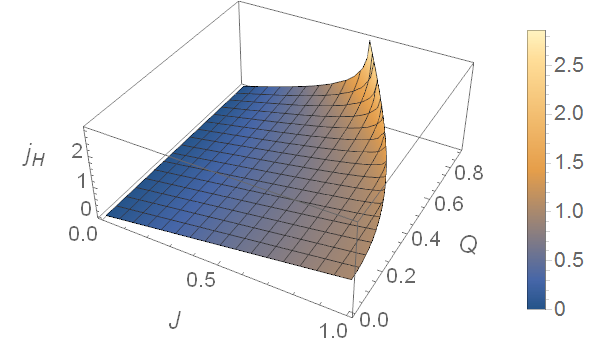}
\caption{}
\label{figureks3}
\end{subfigure}
\ \ \ \ \ \
\begin{subfigure}[b]{0.38\textwidth}
\includegraphics[width=1.0\linewidth]{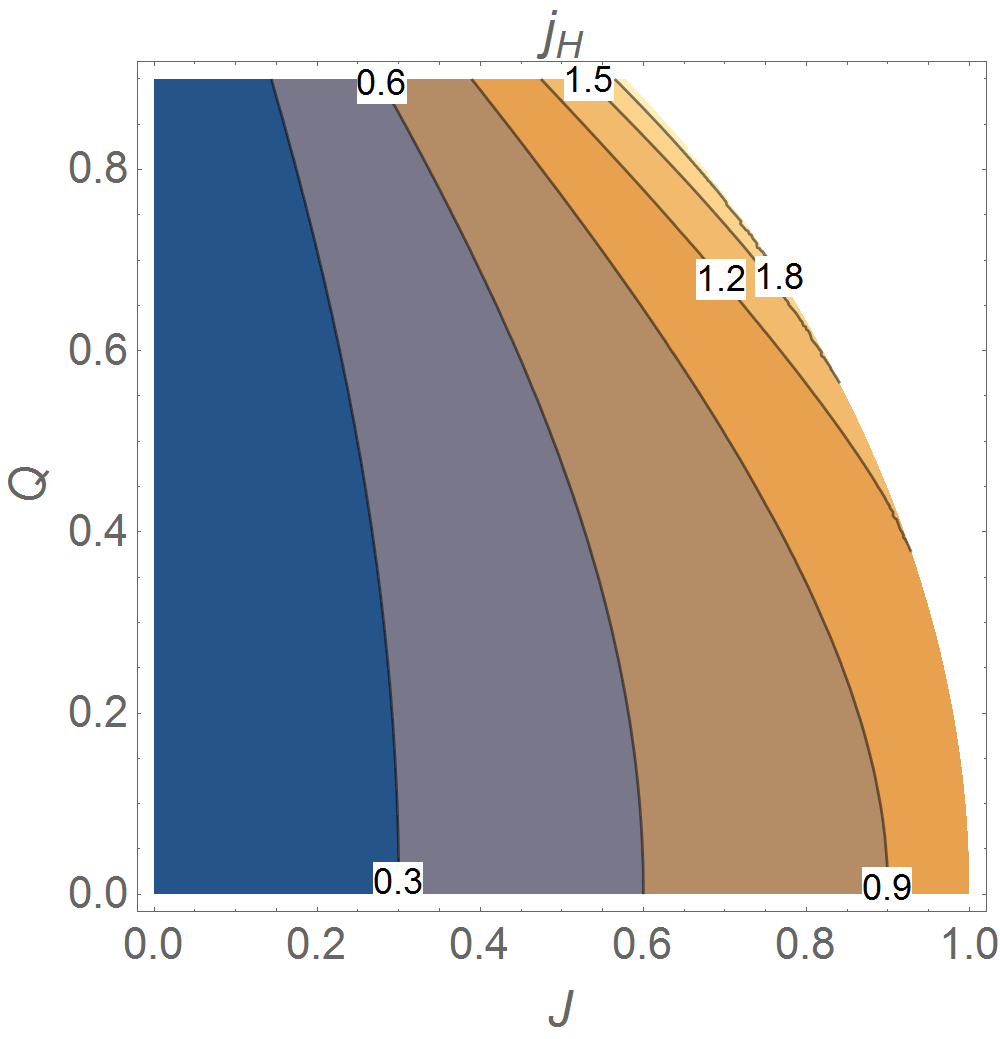}
\caption{}
\label{figureks4}
\end{subfigure}
\caption{
{\small
(a) 3D and (b) 2D contour plots of $j_H\equiv J_{H}/M_{H}^2$ for the Kerr-Sen BH. This quantity becomes larger than unity for sufficiently large charge and angular momentum.}
}
\label{figure4}
\end{figure}

\begin{figure}[H]
\centering

\begin{subfigure}[b]{0.49\textwidth}
\includegraphics[width=1.0\linewidth]{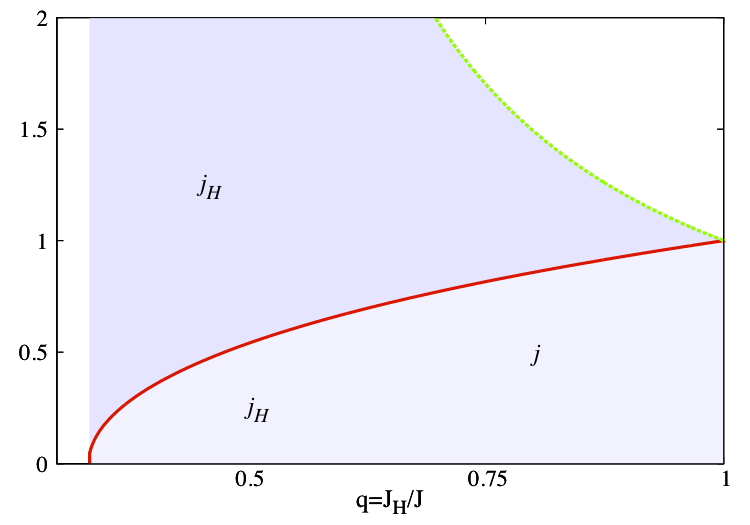}
\caption{}
\label{figurepq1}
\end{subfigure}
\ 
\begin{subfigure}[b]{0.49\textwidth}
\includegraphics[width=1.0\linewidth]{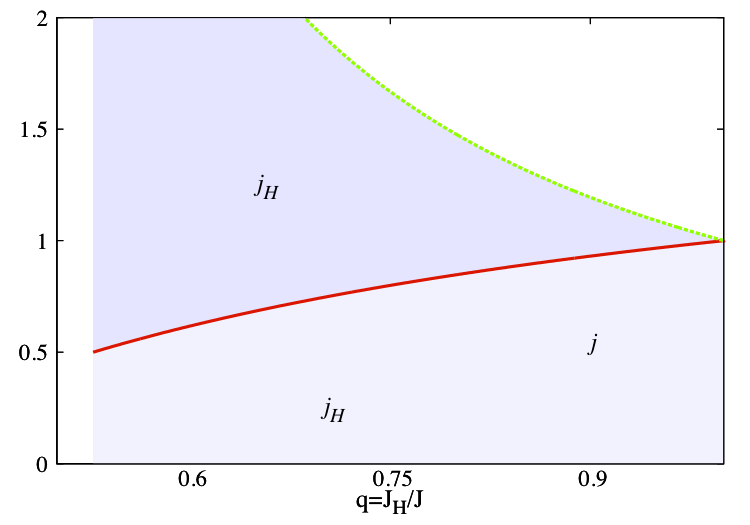}
\caption{}
\label{figurepq2}
\end{subfigure}
\caption{
{\small
Dimensionless ADM, $j$ (light shaded area), and horizon, $j_H$ (light plus dark shaded area) angular momentum, $vs.$ fraction of the angular momentum in the horizon, $q$ for: (a) Kerr-Newman BHs; (b) Kerr-Sen BHs. In both panels, the green dashed (red solid) line corresponds to $j_H$ ($j$) for extremal solutions.}
}
\label{figurepqcharged}
\end{figure}

\begin{figure}[H]
\centering

\begin{subfigure}[b]{0.49\textwidth}
\includegraphics[width=1.0\linewidth]{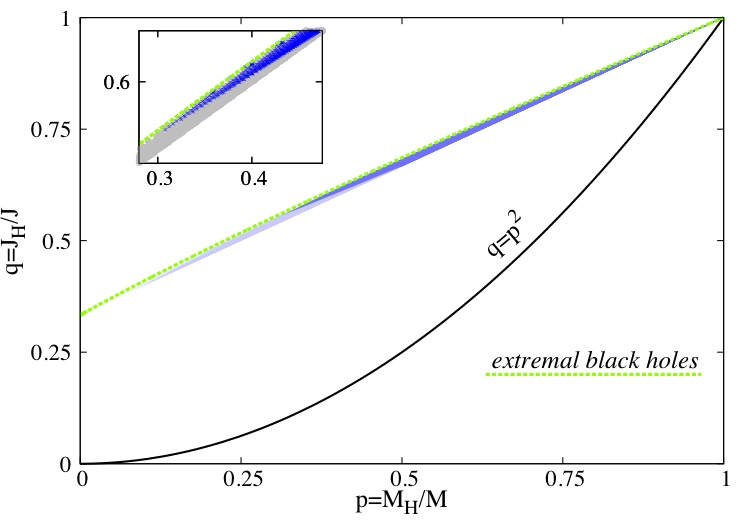}
\label{figurev2}
\end{subfigure}
\caption{
{\small
$p$-$q$ diagrams for Kerr-Newman (light grey) and Kerr-Sen (dark blue) BHs. Unlike the case of Kerr BHs with scalar hair -- Fig.~\ref{figures1} --, \textit{all solutions here} have larger (or equal, in the vacuum Kerr limit) $j_H$ than $j$ and exist along a narrow ribbon on this diagram. The green dashed line corresponds to extremal solutions.}
}
\label{figurepqcharged2}
\end{figure}
Concerning the Kerr bound, we recall that the existence of an event horizon in the Kerr-Newman metric requires $q^2_E+j^2\leqslant 1$, which, in particular implies the Kerr bound $j\leqslant 1$. In terms of the horizon quantities, however, this bound is strongly violated. This is shown in Figs.~\ref{figurekn3}-\ref{figurekn4}. Observe that for $Q=0$, $J_H=J$ and $M_H=M$ and hence the bound is valid. For $Q> 0$, the bound is violated for sufficiently large values of $J$, but which obey $j\leqslant 1$.

Now we turn to the Kerr-Sen BH. As for the Kerr-Newman case let us start by considering the static limit, where one finds a dilatonic charged BH. Firstly, analyzing~\eqref{rhks} we observe that the analogous to the RN bound is now $Q^2/M^2\leqslant 2$ (but which violates the standard RN bound). Moreover, the  extremal limit of the static BH is singular, $i.e$ the areal radius of the horizon vanishes in the extremal limit. Then, from~\eqref{MassHorizonKerrSen} and~\eqref{ChargeHorizonKerrSen}, using also~\eqref{rhks}, we have 
\begin{equation}
M_H =  M \left[ 1 - \frac{ Q^2 }{ 2M^2}  \right] \ , \qquad  
Q_H =  Q \left[1  - \frac{ Q^2 }{2M^2} \right] \ .
\end{equation}
One concludes that both the horizon mass and charge vanish in the extremal limit. This is consistent with the fact that the horizon area also vanishes in this limit. But one also observes the curious feature, which in fact extends to the general Kerr-Sen BH family, as can be seen from~\eqref{MassHorizonKerrSen} and~\eqref{ChargeHorizonKerrSen}, that 
\begin{equation}
\frac{Q_H}{M_H} = \frac{Q}{M}  \ .
\end{equation}
Thus, the Kerr-Sen charge to mass ratio asymptotic bound is not violated by the horizon quantities. 

In Figs.~\ref{figureks1}-\ref{figureks2} and ~\ref{figureks3}-\ref{figureks4} we produce analogous plots for the Kerr-Sen BH to the ones in Figs.~\ref{figurekn1}-\ref{figurekn2} and ~\ref{figurekn3}-\ref{figurekn4} for the Kerr-Newman. The qualitative behaviour of all quantities is similar, but observe that the charge, in the Kerr-Sen case can vary up to $\sqrt{2}$. In particular the Kerr bound can also be violated in terms of horizon quantities -- Fig.~\ref{figureks3}-\ref{figureks4}.

The violations of the Kerr bound in terms of horizon quantities for Kerr-Newman and Kerr-Sen BHs, can be also appreciated in terms of the fraction of horizon angular momentum in Fig.~\ref{figurepqcharged}. One can appreciate a clear similarity in the behaviour of the horizon quantities with that seen for hairy BHs  (Fig.~\ref{figures2}): violations are stronger when a larger fraction of the angular momentum is \textit{outside} the horizon. But in contrast with the case of hairy BHs, the ADM dimensionless angular momentum never exceeds unity.

In Fig.~\ref{figurepqcharged2}, we exhibit the $p$-$q$ diagram for Kerr-Newman and Kerr-Sen BHs. This diagram shows that, in contrast with hairy BHs, the charged BHs always have a larger (or equal in the vacuum Kerr limit) horizon dimensionless angular momentum, as compared to the asymptotic one. In both Kerr-Newman and Kerr-Sen cases, $p$ and $q$ reach one, corresponding to the vacuum limit. But none of this solutions extends to $p,q=0$, since there is no solitonic limit, unlike the case of hairy BHs. 

\begin{figure}[H]
\centering

\begin{subfigure}[b]{0.49\textwidth}
\includegraphics[width=1.0\linewidth]{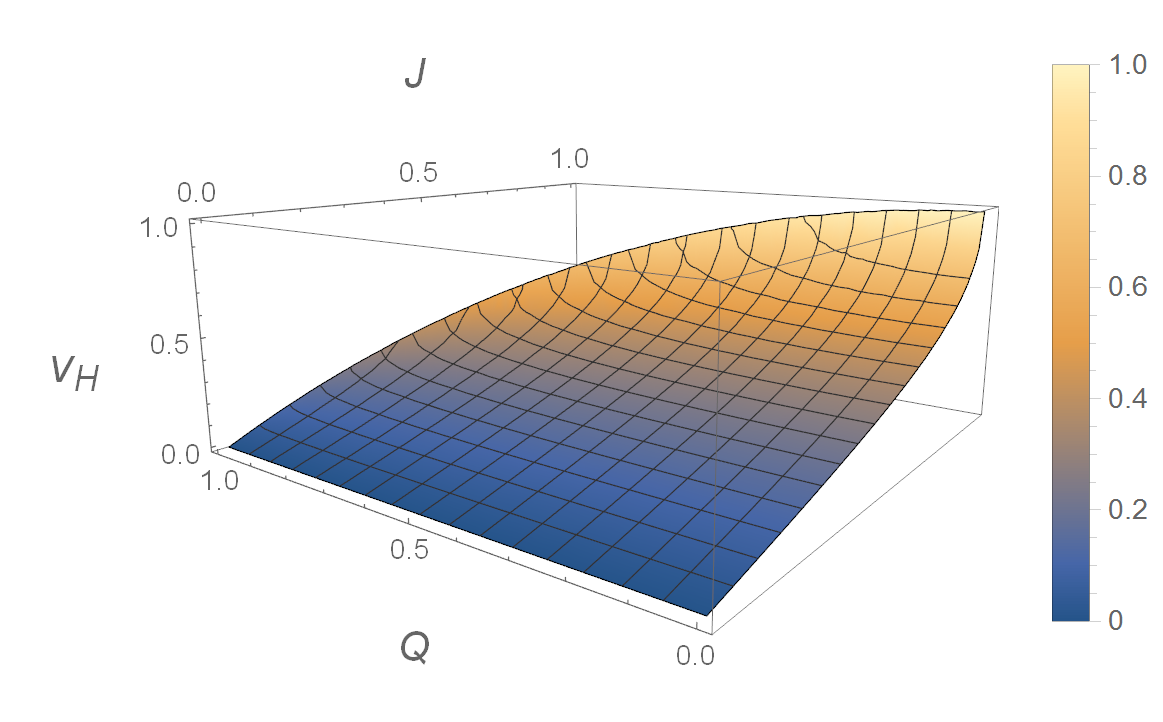}
\caption{}
\label{figurevel1}
\end{subfigure}
\ 
\begin{subfigure}[b]{0.49\textwidth}
\includegraphics[width=1.0\linewidth]{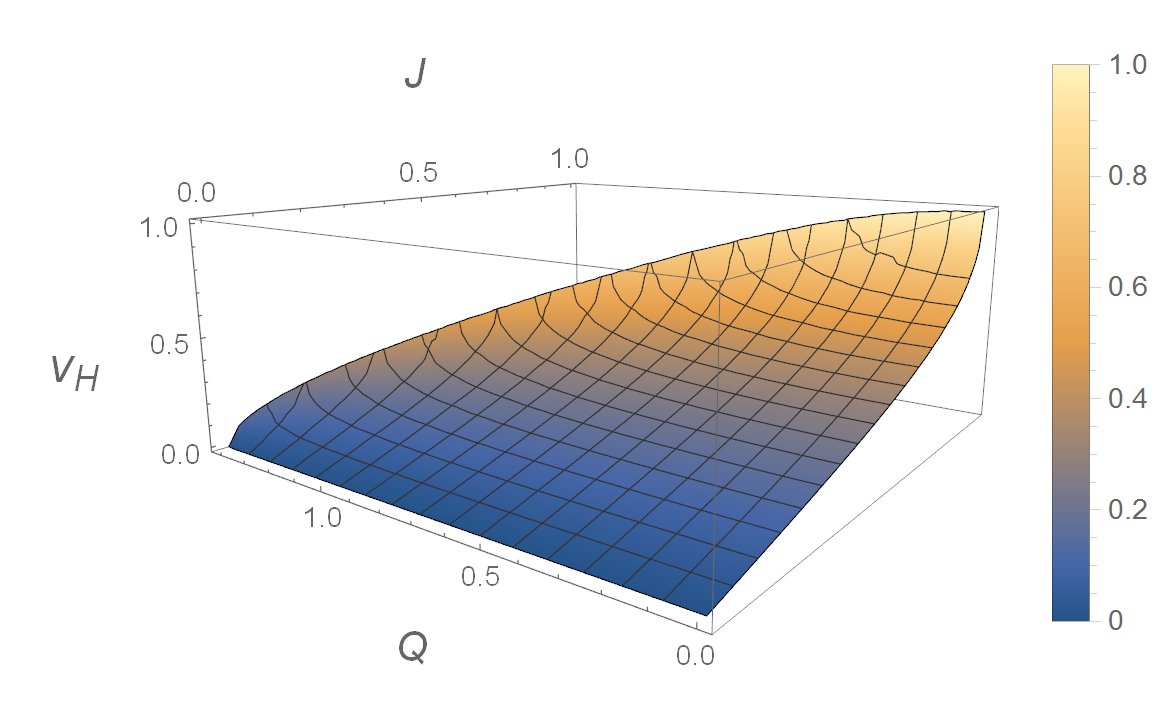}
\caption{}
\label{figurevel2}
\end{subfigure}
\caption{
{\small
Horizon linear velocity for (a) Kerr-Newman and (b) Kerr-Sen BHs. It never exceeds unity (the speed of light).}
}
\label{figurev}
\end{figure}

We now observe that, despite the (unlimited) violations of the Kerr bound there is still a bound on the rotation. It was suggested in~\cite{Herdeiro:2015moa}, that a meaningful linear horizon velocity could be defined in the following way, for a stationary, axisymmetric BH. On the spatial sections of the event horizon one should compute the proper length $L$ of the orbits of the $U(1)$ Killing vector field. Choosing the maximum of such proper lengths, $L_{\rm max}$, that typically occurs at the equator, one defines the circumferencial radius as $R=L_{\rm max}/2\pi$. The horizon linear velocity is simply 
\begin{eqnarray}
v_H=R\Omega_H,
\end{eqnarray}
 where $\Omega_H$ is the angular velocity of the horizon. 
It was then conjectured in~\cite{Herdeiro:2015moa} that, 
for stationary, axisymmetric, asymptotically flat, four dimensional BH solutions,  
$v_H\leqslant c$, where $c$ is the velocity of light, with the equality attained only for vacuum Kerr. 
In~\cite{Herdeiro:2015moa} and~\cite{Herdeiro:2016tmi} we have verified this conjecture for BHs with scalar and Proca hair.
 Here we note that for the Kerr-Newman and Kerr-Sen solutions one can get explicit expressions for $v_H$, which read, in terms of the asymptotic charges:
\begin{eqnarray}
&&
v_H = \frac{J}{M (2M r_H - Q^2)} \sqrt{4M^2 - 2Q^2 + Q^2 \frac{2 J^2 + M^2 Q^2}{J^2 + M^2 Q^2 - 2 M^3 r_H}} \ , \qquad {\rm  Kerr-Newman} \ ,
\\
%
&&
v_H = \frac{J}{M r_H} \sqrt{\frac{ J^2 + M r_H \left(Q^2 - 2M^2 \right)}{J^2 - 2M^3 r_H}} \ , \qquad {\rm 
~~~~~~~~~~~~~~~~~Kerr-Sen} \ , 
\end{eqnarray}
where the corresponding $r_H$ is defined in eq.~\eqref{rhkn} and eq.~\eqref{rhks}. In Fig.~\ref{figurev} we have plotted $v_H$ in terms of the asymptotic electric charge and angular momentum. As can be observed, the linear velocity bound is verified and equality is only attained for vacuum extremal Kerr.

\section{Remarks}
\label{sec_discussion}
In this letter we have pointed out that well known, in closed form, asymptotically flat, charged rotating BH solutions in four spacetime dimensions, violate the Kerr bound in terms of horizon quantities. Furthermore, the Kerr-Newman BH family even violates the RN bound in terms of the horizon quantities, in particular in the static limit, when it reduces to the RN solution. Whereas the computation of these horizon quantities is an exercise that can be found in General Relativity textbooks (see $e.g.$~\cite{PoissonRelativistsToolkit}) the observation that they yield violations of these well-known bounds has not, to the best of our knowledge, been explicitly made. Moreover there is still some widespread belief that these bounds play a fundamental role in BH physics, and whereas counter examples are known, these are often numerical~
\cite{Herdeiro:2014goa,Herdeiro:2016tmi,Petroff:2005vu,Kleihaus:2011tg}
 or exotic solutions~\cite{Herdeiro:2008kq,Herdeiro:2009qy}.
It is thus pedagogical to see that these violations also occur at the level of closed form, non-exotic -- $i.e$ four dimensional, asymptotically flat, regular on and outside a horizon -- solutions, albeit (for the examples herein) for horizon quantities only.

Our motivation for this investigation arose mainly from recent work on Kerr BHs with scalar~\cite{Herdeiro:2014goa} and Proca~\cite{Herdeiro:2016tmi} hair, which are four dimensional, asymptotically flat BH solutions regular on and outside an event horizon and wherein violations of the Kerr bound occur for both ADM and horizon quantities. The former are connected to the existence of a solitonic limit -- scalar boson stars~\cite{Schunck:2003kk} or Proca stars~\cite{Brito:2015pxa} --, and to  the violation of the ADM Kerr bound by those solitonic configurations\footnote{Violations of the Kerr bound in terms of ADM quantities 
were found also for a set of BHs of the dilaton Gauss-Bonnet extension of 
General Relativity \cite{Kleihaus:2011tg,Kleihaus:2015aje}.
In the absence of a solitonic limit in that case, this feature can be attributed to the existence of regions with a negative {\it effective}
energy density \cite{Kleihaus:2015aje}.
}. 
One could then wonder whether the violations in terms of horizon quantities are related to the existence of some matter outside the horizon for these hairy solutions, which is in equilibrium with the BH; this matter, moreover, is an independent quantity (primary hair). In this respect, an  interpretation put forward in~\cite{Herdeiro:2009qy,Herdeiro:2015moa} is that the existence of matter outside the horizon effectively increases the moment of inertia of the BH, since it has to ``drag" a heavy environment, and thus, more dimensionless angular momentum is permitted within the horizon. The BH horizon, however, does not rotate ``too fast", as suggested by the computation of the horizon linear velocity. 

The examples herein show that these violations in terms of horizon quantities can occur even without ``matter" outside the horizon. Indeed, we have observed these violations both for  ``bald" solutions (Kerr-Newman) or solutions that can be interpreted as having secondary (non-independent scalar) hair (Kerr-Sen). Even though there is no matter outside the horizon, there is certainly energy ($e.g.$ electromagnetic). Thus, one may conclude that, in accordance with the elementary principles of Relativity, dragging \textit{energy} like dragging \textit{matter}, takes its toll: it effectively increases the moment of inertia of the rotating BH, and the horizon's ability to carry more specific angular momentum, without rotating too fast.

Finally, we remark that all investigation in this paper is within classical General Relativity. At the quantum level, recent work on BH soft hair, starting with~\cite{Hawking:2016msc}, discussed the existence of horizon conserved charges and possible impact on the BH information loss problem. This reinforces that investigating horizon conserved quantities may lead to new insights into BH physics.

\section*{Acknowledgements}
C. H. and E. R. acknowledge funding from the FCT-IF programme. 
This work was partially supported by  the  H2020-MSCA-RISE-2015 Grant No.  StronGrHEP-690904, and by the CIDMA project UID/MAT/04106/2013. 

\bigskip

\bibliography{bibtex_calculation}
\bibliographystyle{h-physrev4}

\end{document}